# Multiphase flows simulation with the Smoothed Particle Hydrodynamics method

Carlos E. Alvarado-Rodríguez[1], Jaime Klapp[2], J.M. Domínguez[3], A. R. Uribe-Ramírez[1], J.J. Ramírez-Minguela[1], and M. Gómez-Gesteira[3]

[1] División de Ciencias Naturales y Exactas, Universidad de Guanajuato, Col. Noria Alta s/n, C.P. 36050, Guanajuato, Gto., México.
ce.alvarado@ugto.mx
[2] Instituto Nacional de Investigaciones Nucleares (ININ), Km. 36.5 Carretera México-Toluca, La Marquesa, 52750 Estado de México, México.

[3] Environmental Physics Laboratory (EPHYSLAB), University of Vigo, Spain

**Abstract.** This work presents a new multiphase SPH model that includes the shifting algorithm and a variable smoothing length formalism to simulate multiphase flows with accuracy and proper interphase management. The implementation was performed in the DualSPHysics code, and validated for different canonical experiments, such as the single-phase and multiphase Poiseuille and Couette test cases. The method is accurate even for the multiphase case for which two phases are simulated. The shifting algorithm and the variable smoothing length formalism has been applied in the multiphase SPH model to improve the numerical results at the interphase even when it is highly deformed and non-linear effects become important. The obtained accuracy in the validation tests and the good interphase definition in the instability cases, indicate an important improvement in the numerical results compared with single-phase and multiphase models where the shifting algorithm and the variable smoothing length formalism are not applied.

## 1 Introduction

Multiphase flow is an important area of study in many industrial practices such as chemical engineering, environmental analysis and oil recovery processes. Nowadays, there exist two main approaches to analyze this kind of phenomena: laboratory experiments and numerical models. In many cases, laboratory experiments are difficult to perform or very expensive for some of the processes mentioned through this article because of the complex processes involved and the irregular boundary conditions necessary to accurately represent these phenomena.

Computational Fluid Dynamics (CFD) has become a useful tool in science and engineering. The recent advances in computer hardware have increased the applicability, range, and resolution of simulations that CFD methods can properly solve. There are two main approaches to solve the CFD equations, the Eulerian and the Lagrangian formalism. Usually the Eulerian formalism is linked to the use of a mesh to discretize the domain, for example, the finite elements, finite differences, and finite volume methods. These mesh-based methods have been used to simulate multiphase flows but one of their drawbacks is their inability to describe large deformation flows, which can be properly modelled by the meshless Smoothed Particle Hydrodynamics (SPH) method. The SPH method present some advantages comparing with Eulerian methods as: easy to model complex geometries, good representation for mobile boundaries and free surfaces, easy to follow the evolution of the continuous medium, and others.

This method has a simple physical interpretation that allows including sophisticate physics, thus presenting an advantage for new developments. Several SPH multiphase methods have been proposed to simulate flows with different viscosity and density ratios. For example, [1] developed an original SPH model to treat two-dimensional multiphase flows with low density ratios, where the main contribution of their model was a new implementation of the particle evolution according to [2]. Tartakovsky and Meakin (2006) [3] developed a model that combines particle density number based equations and inter-particle forces, this last implementation prevents the presence of artificial surface tension at the interphase between fluids that is a common issue of the standard SPH formulation. Monaghan (1992) [4] proposed a variation in the viscous term of the standard SPH, the modification consists of a small increase in the pressure term when the interaction is among particles from different phases.

Among the available options, due to the advantages with respect to other SPH models, DualSPHysics has been chosen as the platform to implement the multiphase model. DualSPHysics is an open-source code developed at the University of Vigo (Spain) and the University of Manchester (UK) that can be freely downloaded from www.dual.sphysics.org [5]. DualSPHysics is implemented in C++ and CUDA and is designed to perform simulations either on multiple CPU´s or on GPU´s. Due to the SPH method has a high computational cost, the DualSPHysics code presents the advantage to perform the mathematical calculus in a parallel way using all unit processors in a CPU or GPU device, furthermore, the code has tools to perform simulations with irregular boundary conditions easily. The code is performed by duplicate, one version for CPU and the other for GPU devices, this presents some advantages to implement new physical models starting in CPU version and then export the new model to the GPU version.

The implemented optimizations in DualSPHysics allows performing high-resolution simulations in reasonable computational times due to its high level of parallelization in both CPUs and GPUs [6]. The main objective of this work is to present a multiphase model that consists of a combination of different approaches, mainly the shifting algorithm and the variable smoothing length formalism, to improve the treatment for cases with high non-linear deformations at the interphase. Several validation cases are presented, Poiseuille flow and Couette flow, comparing the results provided by the new implementation and the analytical solution. Also three applications cases were conducted, namely, the Rayleigh-Taylor and Kelvin-Helmholtz instabilities, and the oil-water two phase flow in a pipe where the numerical results are compared with the experimental data in the literature.

## 2 The SPH method

In the SPH method, the fluid is discretized as a set of points or fluid elements called particles. The integral equations that govern the fluid dynamics are transformed to a discrete description. This new set of equations is solved in the Lagrangian formalism. The values of the main physical variables (position, velocity, density and pressure) for each particle are obtained as the interpolation of the values of the neighboring particles. A function called kernel is used to transform the continuous medium (fluid) to a discrete description (particles). For a more detailed description of the method the reader is referred to [4], [7], [8], [9], [10].

$$A(\boldsymbol{r}) = \int A(\boldsymbol{r}')W(\boldsymbol{r} - \boldsymbol{r}', h)d\boldsymbol{r}' \quad , \tag{1}$$

where $\boldsymbol{r}$ is the position vector, $W$ the kernel, and $h$ the smoothing length which is a distance larger than the initial particle spacing. The previous equation can be re-written in discrete notation as

$$A(\boldsymbol{r}) = \sum_j m_j \frac{A_j}{\rho_j} W(\boldsymbol{r} - \boldsymbol{r}_j, h) \quad , \tag{2}$$

where the subscript $j$ represents all the neighbouring particles inside the kernel function domain; these are the particles whose contribution cannot be neglected. The derivative of a function can be rewritten in discrete notation as

$$\nabla A(\boldsymbol{r}) = \int A(\boldsymbol{r}')\nabla W(\boldsymbol{r} - \boldsymbol{r}', h)d\boldsymbol{r}' \approx \sum_j m_j \frac{A_j}{\rho_j} \nabla W(\boldsymbol{r} - \boldsymbol{r}_j, h) \quad . \tag{3}$$

The kernel function is a key parameter in the SPH method. There is a wide range of functions that can be used as kernel [11], [12], [13] but all of them must satisfy certain conditions: positivity, compact support and normalization. In addition, $W(r,h)$ must have a delta function behaviour when $h$ goes to zero, and be monotonically decreasing with the distance between particles. For this work we use the kernel function proposed by [13] which can be written as:

$$W(r, h) = \alpha_D \left(1 - \frac{q}{2}\right)^4 (2q + 1) \qquad 0 \le q \le 2 \quad , \tag{4}$$

where $\alpha_D = 7/(4\pi h^2)$ in 2D and $\alpha_D = 21/(16\pi h^2)$ in 3D, $q = r/h$ with r being the distance between particles i and j.
According with [11], this kernel provide a high order of interpolation at a moderate computational cost.

The equations that govern the fluid are presented in the following sections. In section 2.1 the standard SPH formulation implemented in DualSPHysics is introduced. Since the aim of this work is to properly represent multiphase flows, the standard formulation falls short and additional formulations and approximations are required. All changes performed are presented in section 2.2.

### 2.1 The standard SPH model

The momentum and continuity equations used in this work can be expressed as a continuous field as follows:

$$\frac{d\boldsymbol{v}}{dt} = -\frac{1}{\rho}\nabla P + \boldsymbol{g} + \Gamma \quad , \tag{5}$$

$$\frac{d\rho}{dt} = -\rho\nabla\boldsymbol{v} \quad , \tag{6}$$

where $\boldsymbol{v}$ is velocity, $t$ is time, $P$ is pressure, $\rho$ represents the density, $\boldsymbol{g}$ the gravitational acceleration, and $\Gamma$ refers to dissipative terms.

The governing equations in the standard SPH formulation (continuity and momentum) referred to particle $i$ are

$$\frac{d\rho_i}{dt} = \sum_j m_j \boldsymbol{v}_{ij} \nabla_i W_{ij} \quad , \tag{7}$$

$$\frac{d\boldsymbol{v}_i}{dt} = -\sum_j m_j \left(\frac{P_j}{\rho_j{}^2} + \frac{P_i}{\rho_i{}^2} + \Gamma\right) \boldsymbol{\nabla}_j W_{ij} + \boldsymbol{g} \quad , \tag{8}$$

where $P_j$ and $\rho_j$ denote the pressure and density of neighbouring particles, respectively.

The dissipative term $\Gamma$ is implemented in two different ways in DualSPHysics, namely, the artificial viscosity proposed by [4], and the laminar viscosity plus subparticle scale (SPS) turbulence [14], [15], [34].

Artificial viscosity is frequently used due to its stability and simplicity. The viscosity term $\Gamma$ in equation (8) can be written for artificial viscosity as:

$$\Gamma = \begin{cases} \frac{-\alpha\bar{c}_{ij}\mu_{ij}}{\rho_{ij}} & \boldsymbol{v}_{ij} \cdot \boldsymbol{r}_{ij} < 0 \\ 0 & \boldsymbol{v}_{ij} \cdot \boldsymbol{r}_{ij} > 0 \end{cases} \quad , \tag{9}$$

where $\boldsymbol{r}_{ij} = \boldsymbol{r}_i - \boldsymbol{r}_j$, $\boldsymbol{v}_{ij} = \boldsymbol{v}_i - \boldsymbol{v}_j$, $\mu_{ij} = h\boldsymbol{v}_{ij} \cdot \boldsymbol{r}_{ij}/(\boldsymbol{r}^2{}_{ij} + \eta^2)$, $\bar{c}_{ij} = 0.5(c_i + c_j)$ is the mean speed of sound, $\eta^2 = 0.01h^2$, and $\alpha$ is a free parameter that must be tuned depending on the problem configuration.

When the laminar + SPS turbulence is used to represent the viscous stresses, the momentum equation (8) can be expressed as:

$$\frac{d\boldsymbol{v}_i}{dt} = -\sum_j m_j \left(\frac{P_j}{\rho_j{}^2} + \frac{P_i}{\rho_i{}^2}\right) \boldsymbol{\nabla}_j W_{ij} + \boldsymbol{g} + \sum_j m_j \left(\frac{4\upsilon_0 \boldsymbol{r}_{ij} \cdot \nabla_i W_{ij}}{(\rho_i + \rho_j)(\boldsymbol{r}^2{}_{ij} + \eta^2)}\right) \boldsymbol{v}_{ij} \quad , \tag{10}$$

where $v_0$ is kinetic viscosity ($10^{-6}$ m²s in the case of water).

Laminar viscosity was used for the validation cases and artificial viscosity for the instability cases, the different selection helps to adequately represent non-linear effects at the interphase in the application cases. These non-linear effects could be smoothed or disturbed by the viscosity laminar treatment.

In the SPH formalism the fluid is considered as weakly incompressible and pressure is calculated as a function of density. Following [17], Tait's equation is used to relate pressure and density. This equation provides high pressure variations at small density oscillations and is written in the form

$$P = B\left[\left(\frac{\rho}{\rho_0}\right)^{\gamma} - 1\right] \quad , \tag{11}$$

where $B = c_0{}^2 \rho_0 / \gamma$, $\rho_0$ is the reference density, $\gamma$ is the polytropic constant which is set to 7, and $c_0 = c(\rho_0)$ the speed of sound at the reference density. $B$ also provides a limit for the maximum change that the density can experience. The speed of sound ($c_0$) is an artificial value that must be, at least, 10 times bigger than the highest fluid velocity estimated for the physical problem under study. This condition only allows a density oscillation of 1% around the reference density ($\rho_0$).

## 2.2 The multiphase SPH model

Different approaches and methods have been proposed to simulate multiphase flows [12], [18], [3]. In this work several new approaches are added to the standard formalism. These features permit to properly simulate multiphase flows, where the main contribution lies in the improved management to interphase with highly non-linear deformations.

### 2.2.1 The momentum equation for the multiphase model

The instability and artificial surface tension produced in a multiphase flow using the standard SPH has been reported by [1] and [19]. For this work we have replaced the equation (8) used in the standard SPH formulation by the expression (12) which to permit that higher density ratios in simulations avoiding the artificial surface tension [1]. The method is rather robust, even for large free-surface fragmentation and folding, efficient and relatively easy-to-code and results stable and capable to easily treat a variety of density ratios [1].

$$\frac{dv_i}{dt} = -\sum_j m_j \left(\frac{P_i + P_j}{\rho_i \rho_j} + \Gamma\right) \nabla_i W_{ij} + \boldsymbol{g} \quad . \tag{12}$$

Higher density ratios can be simulated with the use of this expression avoiding the artificial surface tension.

*2.2.2 The equation of the state for the multiphase model*

According to [1] the pressure of each phase is calculated using the equation of state (11), which is calculated using appropriate parameters according to each phase reference density. So, the equation of state (11) is calculated for each phase using:

$$P_H = B_H \left[ \left( \frac{\rho}{\rho_{0_H}} \right)^{\gamma_H} - 1 \right], \qquad P_L = B_L \left[ \left( \frac{\rho}{\rho_{0_L}} \right)^{\gamma_L} - 1 \right], \qquad (13)$$

where the subscripts $H$ and $L$ denote the fluid with higher and lower density, respectively.

The constant $B_H$ is chosen to permit a small compressibility of the higher density fluid, that is $v_{maxH}/c_H \ll 1$, where $v_{maxH}$ is the maximum velocity of the fluid with higher density expected in the considered problem. Then $B_L$ is matched to $B_H$ in the equation of state for the fluid with lower density to create a stable pressure at the interphase. Moreover this formalism ensures that the fluid stays at rest when $\rho = \rho_H$ or $\rho = \rho_L$ and the pressure is zero. This formulation allows the simulation of high density ratios (e.g. 1:1000, which is similar to the air-water ratio). For the water-air interaction typical values are $\gamma_H = 7$ and $\gamma_L = 1.4$. However, only simulations with lower densities ratios (1:2) and a value of $\gamma = 7$ are considered for the cases presented in this work.

*2.2.3 The shifting algorithm*

The shifting algorithm, henceforth *shifting*, is a new implementation in DualSPHysics. This algorithm was proposed by [20] and it is used to keep a better distribution of particles. This algorithm shifts the position of particles slightly after their normal interaction, due to pressure and velocity and is applied in this work to prevent voids in the particle distribution formed by the interaction between particles with different densities. The magnitude and direction of the position shift is governed by Fick's law, which slightly moves particles from higher to lower particle concentration regions. The displacement is calculated assuming that the flux of particles is proportional to the velocity. So, according to Fick´s law the shifting displacement of a particle can be written as:

$$\delta r_s = -K \nabla C \quad , \qquad (14)$$

where $K = -2$ is considered and the shift, $\delta r_s$, is added to the equation of particle displacement as follow

$$\boldsymbol{r}_i^{n+1} = \boldsymbol{r}_i^n + \Delta t \boldsymbol{v}_i^n + 0.5 \Delta t^2 \boldsymbol{F}_i^n + \delta r_s. \qquad (15)$$

Then, the particle concentration is calculated at each time step from the summation of the kernel function and the concentration gradient in the usual way by

$$C_i = \sum_j V_j W_{ij} \quad , \qquad \nabla C_i = \sum_j V_j \nabla W_{ij} \quad , \qquad (16)$$

where $V_j$ means particle volume and $C_i$ is the concentration of the neighbour particles.

*2.2.4 The variable smoothing length formalism*

The variable smoothing length formalism, henceforth $h_{var}$, was proposed by [21] and [22] to properly describe shock waves and the interphase between two fluids where the density changes by a significant amount. The basic idea is to allow the smoothing length to change from particle to particle through a series of kernels. The $h_{var}$ is calculated from an initial density ($\hat{\rho}$) and smoothing length ($h_0$) as:

$$\hat{\rho}_i = \sum_{j=1}^{N\_n} m_j W\left(\left|r_i - r_j\right|, h_0\right) \quad , \tag{17}$$

where N_n means the number of neighbors.

Then local bandwidth factors, $\lambda_i$, are constructed according to

$$\lambda_i = k \left(\frac{\hat{\rho}_i}{\bar{g}}\right)^{-\epsilon} \quad , \tag{18}$$

$$\log \bar{g} = \frac{1}{N} \sum_{j=1}^{N\_n} log\,\hat{\rho}_j \quad , \tag{19}$$

where k is a constant (k≈1), and $\epsilon$ is a sensitive parameter that ranges from 0 to 1. Then, the $h_{var}$ is calculated according to

$$h_{var} = \lambda_i h_0 \quad . \tag{20}$$

The kernel is symmetrized to conserve linear momentum using the following average for each pair of particles

$$h_{varij} = \frac{h_{vari} + h_{varj}}{2} \quad . \tag{21}$$

So, the kernel with the new smoothing length will replace the previous version:

$$W_{ij} = W\left(\left|r_i - r_j\right|, h_{varij}\right) \quad . \tag{22}$$

## 2.3 Parallel structure in the code

The DualsSPHysics code presents a parallel structure using the OpenMP and CUDA tools. The code is write by duplicate, one version is write in the C++ computational language for CPU processors where all unit processor are used by default using OpenMP and other version is write in CUDA computational language for use the internal CUDA cores in one GPU processor.

In the SPH method the main computational time for simulations is during the particle interaction, so the calculus in the interaction is processed in parallel, then the results are processed in a serial way. For the GPU version, only the interaction are calculated in the CUDA cores and the preprocessing and post-processing are handled by the CPU processor. The mean characteristics of the DualSPHysics code can be consulted in [5]. The advantage of the use parallel codes are describe in [23], [24].

## 3 Validation

Several improvements to the numerical method have been mentioned in previous sections. All of them focused on the multiphase treatment and aimed to increase the accuracy of the model. This section provides four test cases that highlight the accuracy of our multiphase model implemented in DualSPHysics. The case 3.1 evidences the accuracy in the evolution of the velocity profile within a duct for the Poiseuille flow test case with two densities. The case 3.2 shows the accuracy in the evolution of the velocity profile when the velocity of the fluid is induce by a boundary simulating a Couette flow with two densities. The results from SPH simulations are compared with the analytical solution for all cases.

### 3.1 Poiseuille flow with two densities

The plane Poiseuille flow consists of a laminar flow produced by a constant pressure gradient between two parallel infinite plates. The plane Poiseuille flow test on SPH has already been performed by [25], [26] and [27]. This case tests out the accuracy of the evolution in the velocity profile when the numerical results are compared with the analytical solution. In this case a Plane Poiseuille flow is conducted for two fluids with different densities and viscosities. The test is simulated in the XY plane (2-D), neglecting the gravitational acceleration, the distance between the plates is 1 mm. The test was performed using periodic conditions in X direction at $x = -0.5\ mm$ and $x = 0.5\ mm$. The laminar viscosity model is used to calculate de momentum equation, but as mentioned before with different density and viscosity values. The initial condition of simulation is shown in Figure 1 and the set-up configuration is summarized in Table 1.

**Table 1.** Set-up configuration for the Poiseuille flow test case with two densities.

| Parameter | Value |
|---|---|
| Total particles | 700; 1,188; 2,600 |
| Initial inter-particle spacing | $4 \times 10^{-5}$, $3 \times 10^{-5}$, $2 \times 10^{-5}$ m |
| Lower density ($\rho^I$) | 500 kg/m$^3$ |
| Higher density ($\rho^{II}$) | 1000 kg/m$^3$ |
| Body force parallel to X-axis ($\textbf{F}$) | $10^{-4}$ m/s$^2$ |
| Viscosity for lower density ($\mu^I$) | $0.5 \times 10^{-6}$ m$^2$/s |
| Viscosity for higher density ($\mu^{II}$) | $1 \times 10^{-6}$ m$^2$/s |

This case evidences the accuracy of the evolution in the velocity profile in multiphase simulations using de multiphase model coupling $h_{var}$ and *shifting*, where the accuracy of the SPH simulations is determined by comparing the numerical and analytical solutions.

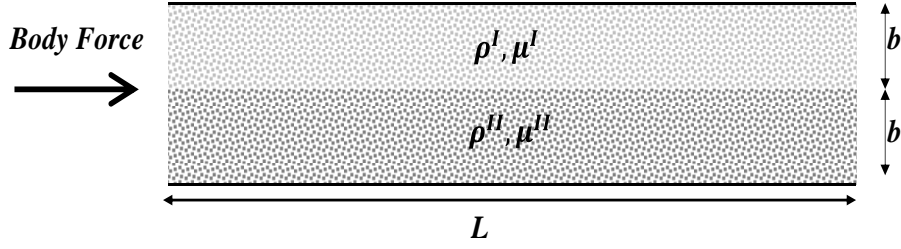

**Fig. 1**. Initial condition for the Poiseuille test case with two densities.

The analytical solution for the steady state was presented in [28] and compared here with the SPH simulation results. The pressure difference used in the analytical solutions is $\Delta P = 1.83x10^{-7} Pa$ . The analytical solutions are

$$\boldsymbol{v}_x{}^I = \frac{\Delta P b^2}{2\mu^I L}\left[\left(\frac{2\mu^I}{\mu^I+\mu^{II}}\right) + \left(\frac{\mu^I-\mu^{II}}{\mu^I+\mu^{II}}\right)\left(\frac{y}{b}\right) - \left(\frac{y}{b}\right)^2\right] \ , \tag{23}$$

$$\boldsymbol{v}_x{}^{II} = \frac{\Delta P b^2}{2\mu^{II} L}\left[\left(\frac{2\mu^{II}}{\mu^I+\mu^{II}}\right) + \left(\frac{\mu^I-\mu^{II}}{\mu^I+\mu^{II}}\right)\left(\frac{y}{b}\right) - \left(\frac{y}{b}\right)^2\right] \ , \tag{24}$$

where $b = 0.5\ mm$ is the height of each fluid, $\mu^I$ and $\mu^{II}$ are the viscosities of fluids I and II, respectively, and $L = 1\ mm$ the length of the container.

Figure 2 shows the velocity ($\boldsymbol{v}_x$) profile at 0.5 seconds once the steady state has been attained for three different resolutions. Velocity was calculated at 21 points located at the same X position and varying the Y position every 0.05 mm.

Numerical and analytical results show good agreement. The convergence test was carried out using three different resolutions corresponding to 700; 1,188 and 2,600 particles, for the same case. The relative error was calculated using equation (25).

$$\% RMSE = 100 \times \sqrt{\frac{1}{N_d}\sum\left(u_{\text{SPH}} - u_{\text{analytical}}\right)^2} \ , \tag{25}$$

where $\% RMSE$ is the relative mean square error, $u_{SPH}$ the SPH velocity, $u_{\text{analytical}}$ the analytical velocity and $N_d$ is the number of data points.

The velocity profile corresponding to the steady state shown in Figure 2 is asymmetric due to the different viscosity values. In this case the top velocity value is located at the low viscosity part of the fluid. These results are in good agreement with theory since the stress tensor decreases when the viscosity decreases.

Results of the convergence test are shown in Table 2 where the relative error (%RMSE) decreases when the resolution increases.

### 3.2 Couette flow with two densities

The second validation case is a Couette flow. Numerical SPH calculations of Couette flow has also been performed in [27] and [25]. This test consists of a laminar flow between two parallel infinite plates produced by the displacement of the top plate with constant velocity. As in previous cases, the infinite plates are represented using periodic boundary conditions and the case is simulated in XY plane (2-D). In this case a Couette flow validation is conducted using two fluids with different densities and viscosities. The set-up configuration is summarized in Table 3 and the initial configuration is depicted in Figure 3. This case tests the accuracy of the evolution in the velocity profile when the motion of the fluid is induced by a boundary and the simulation is performed with two densities.

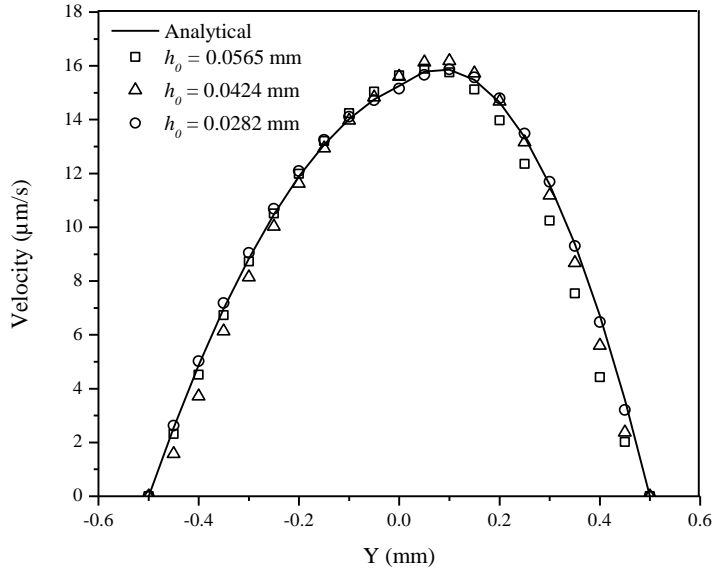

**Fig. 2.** Numerical and analytical solution for the Poiseuille test case with two densities at t = 0.5 seconds, when the steady state has been attained.

**Table 2.** Numerical parameters and errors for SPH velocity profiles shown in Figure 2.

| Np | Dp (mm) | $h_0$ (mm) | %$RMSE$ at the steady state |
|---|---|---|---|
| 700 | 0.04 | 0.0565 | $8.44 \times 10^{-5}$ |
| 1,188 | 0.03 | 0.0424 | $6.01 \times 10^{-5}$ |
| 2,600 | 0.02 | 0.0282 | $3.55 \times 10^{-5}$ |

**Table 3.** Set-up configuration for the Couette flow test case with two-densities.

| Parameter | Value |
|---|---|
| Total particles | 2,277; 5,976; 12,948 |
| Initial inter-particle spacing | 5x10⁻⁵, 4x10⁻⁵, 2x10⁻⁵ m |
| Lower density ($\rho_1$) | 1000kg/m³ |
| Higher density ($\rho_2$) | 2000kg/m³ |
| Viscosity for lower density ($\mu_1$) | 0.5x10⁻⁶m²/s |
| Viscosity for higher density ($\mu_2$) | 1x10⁻⁶m²/s |

The top plate moves with a constant velocity $V_p = 1x10^{-3} m/s$ in X-direction. The analytical solution for a steady state was presented in [29] and compared here with the results provided by SPH simulations. The analytical solution is

$$v_{\rho 1} = \frac{\mu_1 V_p}{\mu_2 b_1 + \mu_1 b_2} y \quad , \qquad (26)$$

$$v_{\rho 2} = \frac{V_p}{\mu_2 b_1 + \mu_1 b_2} (\mu_1 (y - b_1) + \mu_2 b_1) \quad , \qquad (27)$$

where $b_1 = b_2 = 0.5$ mm and $\mu_1$ and $\mu_2$ are indicated in Table 3.

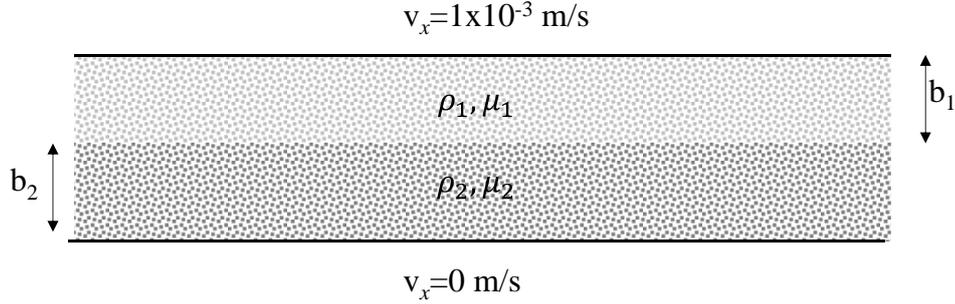

**Fig. 3.** Initial configuration for the Couette flow with two densities.

Figure 4 shows the comparison between SPH calculations and the analytical solution for the Couette flow with two densities. In this case, the velocity profile at the steady state is not linear as in Figure 4 due to the different viscosity and density values.

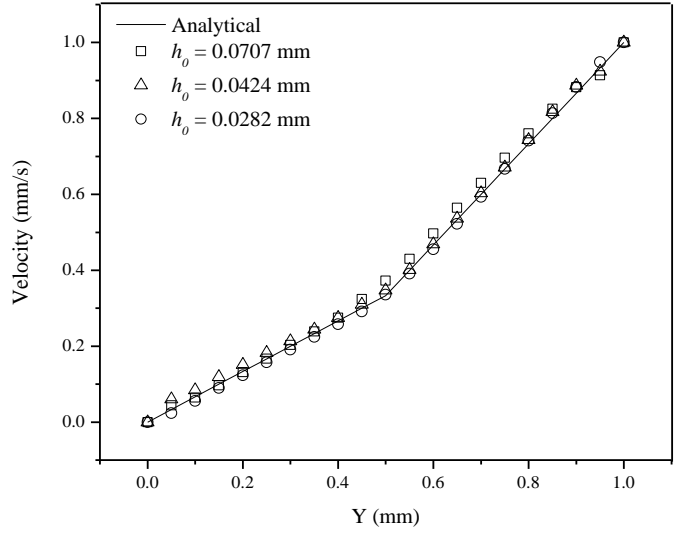

**Fig. 4**. Numerical and analytical solutions for the Couette test case with two densities.

The agreement between numerical results and the analytical solution for the Couette flow with two densities is good and the accuracy increases with the resolution. The relative error was calculated using equation (25). Results from the convergence test are shown in Table 4.

**Table 4.** Numerical parameters and errors for SPH velocity profiles shown in Figure 4.

| Np | Dp (mm) | $h_0$ (mm) | %RMSE at the steady state |
|---|---|---|---|
| 2,277 | 0.05 | 0.0707 | $10^{-6}$ |
| 5,976 | 0.04 | 0.0424 | $1.31 \times 10^{-6}$ |
| 12,948 | 0.02 | 0.0282 | $9.63 \times 10^{-7}$ |

The results of the cases 3.1 and 3.2 prove the good accuracy in multiphase simulations using multiphase model coupling $h_{var}$ and *shifting*.

### 3.3 Rayleigh-Taylor instability

The Rayleigh-Taylor instability was numerically studied in [30], [4] and [31]. This test case presents two different challenges, namely, the proper reproduction of the interphase between two different fluids and the reproduction of non–linear effects involved in this evolution case. Thus the Rayleigh-Taylor instability is a perfect case to test the improvement of coupling $h_{var}$ and *shifting* in the multiphase model.

The initial set-up is described here, two fluids are confined in a rectangular container and the interphase between them is set to be at $y = 0.15\sin(2\pi x)$ to create an initial perturbation. In order to be consistent with [30], the gravity acceleration is -1.0 m/s². The set-up configuration is summarized in Table 5 and shown in Figure 5.

**Table 5.** Set-up configuration for the Rayleigh-Taylor instability test case.

| Parameter | Value |
|---|---|
| Total particles (Np) | 20,901; 81,801; 232,601 |
| Initial inter-particle spacing (Dp) | 10.0, 5.0, 0.9 mm |
| Lower density ($\rho_1$) | 1,000 kg/m³ |
| Higher density ($\rho_2$) | 1,800 kg/m³ |
| Artificial viscosity for lower density ($\alpha_1$) | 0.05 |
| Artificial viscosity for higher density ($\alpha_2$) | 0.05 |

A preliminary simulation was carried out with standard SPH, a second simulation was conducted with the multiphase model coupling $h_{var}$ and *shifting*. Then, two numerical results were compared. Figure 5 presents the same instant (t = 5 seconds) for both simulations.

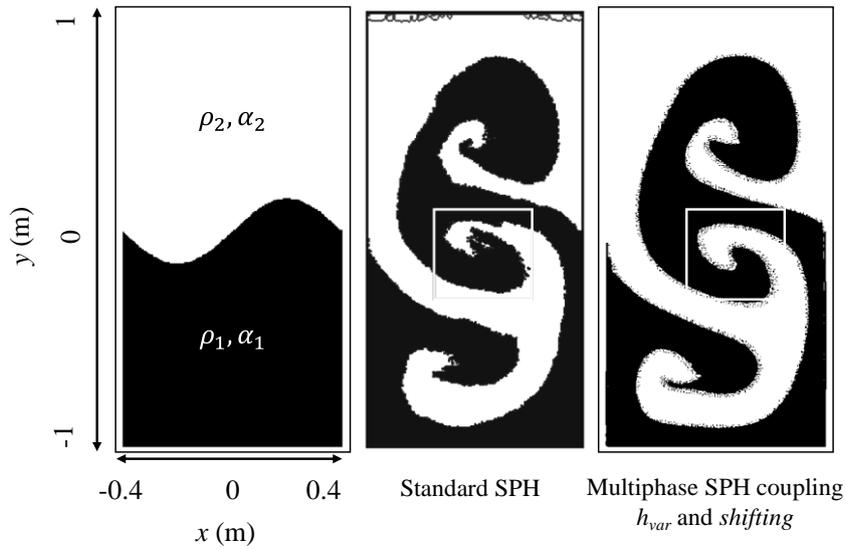

**Fig. 5**. Rayleigh-Taylor instability simulation. Left: Initial conditions. Middle: simulation with standard SPH. Right: Simulation with multiphase model coupling $h_{var}$ and *shifting*.

The general evolution of the system is similar in the both cases, however the definition and shape of the interphase is even better when the $h_{var}$ is coupling with *shifting*, also the void that appear at the top of the simulation (second panel in Figure 5) are prevented. These simulations show how the SPH model can handle nonlinear effects and provide a proper interphase representation. The multiphase model using $h_{var}$ and *shifting* presents better results than those reported by [30] for the Weakly Compressible SPH method and the SPH projection method, in both cases the instability is not totally formed. The reference [4] also reported the Rayleigh-Taylor instability using a simple SPH algorithm for fluids with high density ratios obtaining similar results. However, applying $h_{var}$ and *shifting*, the interphase provides a better definition in zones with high deformation. The reference [31] reported a multiphase model where pressure is continuous at the interphase obtaining similar results.

### 3.4 Kelvin-Helmholtz instability

The Kelvin-Helmholtz instability is a good test to probe that the implemented model can simulate instabilities created by the interaction between two fluids, more precisely the shear stress at the interphase. The test was performed in two dimensions using periodic conditions in X direction at $x = -0.5\ cm$ and $x = 0.5\ cm$ and limited by dynamic boundary layers proposed by [32] at $y = -0.125\ cm$ and $y = 0.125\ cm$. In this case, the instability was simulated using artificial viscosity applying $h_{var}$ and *shifting*. The set-up configuration is presented in Table 6.

**Table 6.** Set-up configuration for the Kelvin-Helmholtz instability test case.

| Parameter | Value |
|---|---|
| Total particles | 501,501 |
| Initial inter-particle spacing | 1.0 mm |
| Lower density ($\rho_1$) | 1,000kg/m³ |
| Higher density ($\rho_2$) | 2,000kg/m³ |
| Artificial viscosity for lower density ($\alpha_1$) | 0.05 |
| Artificial viscosity for higher density ($\alpha_2$) | 0.1 |

The initial velocities in the X direction are 0.5 m/s and -0.5m/s for $\rho_1$ and $\rho_2$, respectively. An initial perturbation at the interphase was set-up using an initial small velocity in the Y direction, $v_y = 0.025\ sin(-2\pi(x + 0.5)/\lambda)$, where $\lambda = 1/6$. The initial conditions of the Kelvin-Helmholtz instability test are shown in Figure 6

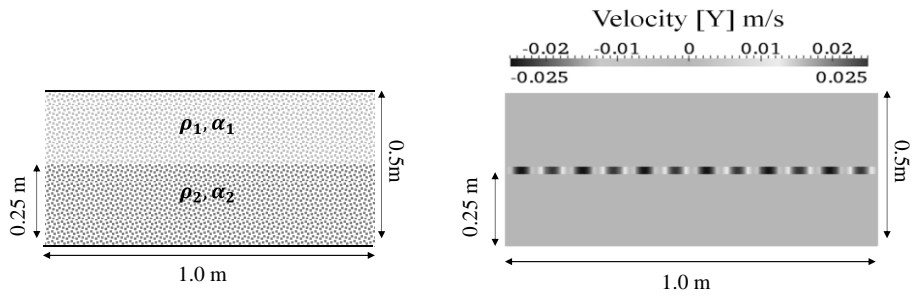

**Figure 6.** Initial condition of the Kelvin-Helmholtz instability test. The left panel shows the positions of two fluids with different densities. The right panel shows the initial perturbation in velocity ($v_y$).

The Kelvin-Helmholtz instability is reproduced using artificial viscosity in standard SPH and the multiphase model, as is shown in the Figure 7. However, results show particle voids and a strait shape at the interface in simulations with standard

SPH. The results improve when $h_{var}$ and *shifting* is applied, which prevents the formation of voids observed in previous simulations and the shape is continuous. The characteristic growth timescale of the incompressible Kelvin-Helmholtz improve of the shape at the interface can be seen better with a zoom in each eddy of the instability as observed in Figure 8. Accordingly, multiphase simulations between two fluids is improved when both $h_{var}$ and *shifting* are applied,

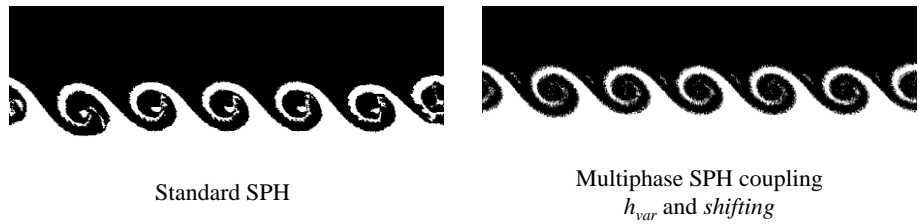

Standard SPH

Multiphase SPH coupling
$h_{var}$ and *shifting*

**Fig. 7**. Results of the Kelvin-Helmholtz instability test.

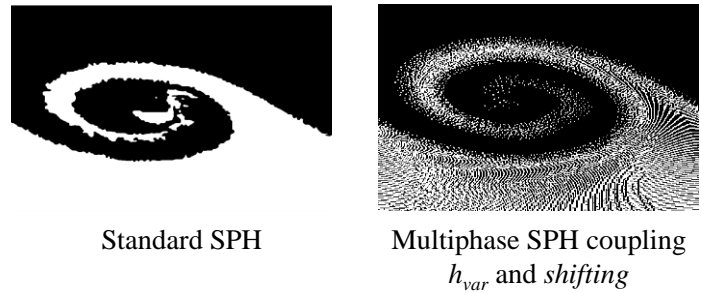

Standard SPH

Multiphase SPH coupling
$h_{var}$ and *shifting*

**Fig. 8**. Comparison at the interface: a) standard SPH, and b) multiphase SPH coupling $h_{var}$ and *shifting*.

Similar results are reported by [33] using a different model for the variable smoothing length and artificial dissipation terms to treat the interphase through the evolution of the discontinuity. Agertz et al., (2007) [34] reported that the standard SPH formulation is not capable of properly modelling dynamical instabilities due to low density SPH particles close to high density regions where particles suffer erroneous pressure forces due to the asymmetric density within the smoothing kernel. Comparing the numerical results reported by [33] and [34] with the results obtained with the model presented in this work, the shape of interface between fluids is improve and the internal vortex is formed clearly and in a continuous way. However, the numerical results should be review comparing with an analytical solution. Therefore, in order to establish an analytical value to validate the numerical results, the characteristic onset time of the Instability in the linear regime given by equation (28) was used to qualitatively compare the numerical values obtained with SPH

$$\tau = \frac{(\rho_1 + \rho_2)\lambda}{\sqrt{\rho_1 \rho_2}|v_2 - v_1|} \tag{28}$$

For the above initial condition values the characteristic time is $\tau = 0.35$ s for $\lambda=1/6$ m, and the numerical results gives e characteristic onset time $\tau = 0.375$.

### 3.5 Oil-water two phase flow in a pipe.

This test shows a very common application in engineering. The simulation is based following the experimental data reported by [35]. This test proves that the multiphase model coupling $h_{var}$ and shifting can simulate the correct flow patterns of two phase oil-water flow in a horizontal pipe. The case was performed in 2D, water and oil are mixed together via a 45° T junction placed at the inlet of the pipe. The overall length and internal diameter of the test section are 2 m and 20 mm, respectively. At first, the flow conditions were determined and oil–water flow was allowed to reach equilibrium. This equilibrium was determined when the fluid velocities are constant, at this time the flow patterns are considered.

Four cases were performed with different inlet values to obtain diverse internal flow patterns in the pipe.

The inlet zone was performed with a constant Poiseuille velocity profile according to the equation (29) where $v_i$ is the inlet velocity for each fluid. Inlet velocities of oil are 0.085, 0.25, 0.085 and 0.65 m/s for case 1, 2, 3 and 4 respectively. Inlet velocities of water are 0.16, 0.20, 1.0 and 0.16 m/s for case 1, 2, 3 and 4 respectively. The initial parameters used in the simulations are shown in the Table 12. The geometry and initial conditions are shown in the Figure 9.

$$\boldsymbol{v}_{inlet} = \boldsymbol{v}_{o(w)}\left(1 - \left(\frac{r}{R}\right)^4\right). \tag{29}$$

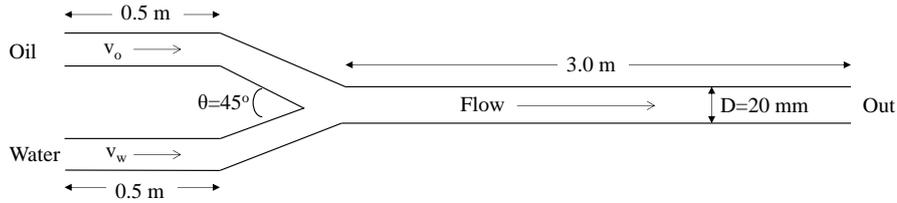

**Fig. 9.** Geometry and dimensions of the water-oil two phase flow in a pipe case. Water and oil are mixed together via a 45° T junction placed at the inlet of the pipe. The $\mathbf{v}_o$ and $\mathbf{v}_w$ are the inlet velocities of oil and water respectively.

**Table 7.** Set-up configuration for the water-oil two phase flow in a pipe.

| Parameter | Value |
|---|---|
| Total particles | 124,533 |
| Initial inter-particle spacing | 0.65 mm |
| Oil density ($\rho_1$) | 840 kg/m$^3$ |
| Water density ($\rho_2$) | 998 kg/m$^3$ |
| Artificial viscosity for oil ($\alpha_1$) | 0.045 |
| Artificial viscosity for water ($\alpha_2$) | 0.01 |

The relation between superficial velocity and real velocity is $u_{sq} = u_q * \alpha_q$, where $u_{sq}$ is superficial velocity of phase q, $u_q$ is real velocity of phase q, and $\alpha_q$ is void fraction of phase q. Void fraction of phase q is $\alpha_q = A_q/A_{tot}$ where $A_q$ is area of phase q, and $A_{tot}$ is total area in a cross-section of the pipe. Generally, seven different types of flow patterns were observed at horizontal pipe, namely; bubbly, slug, smooth stratified, wavy stratified, churn, annular and dual continuous flow. In this work only the bubbly, smooth stratified, wavy stratified and dual continuous flow are simulated and compared visually with the captured photos of experimental flow patterns in a horizontal pipe reported in [35]. In the Figure 10 are shown the patterns got by SPH simulations of the cases 1 to 4.

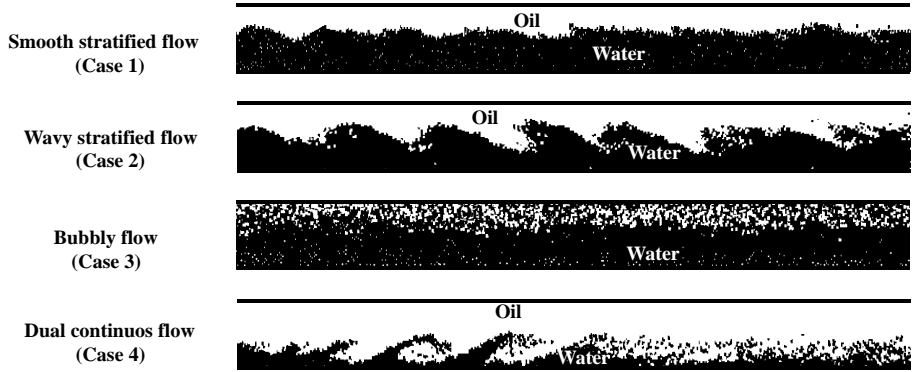

**Fig. 10.** Flow patterns got by SPH simulations using the multiphase model coupling the h$_{var}$ and shifting.

At low inlet oil and water velocities, smooth stratified flow was identified in numerical results. If then inlet water velocity increases, wavy stratified flow was shaped. At high inlet water velocities stratified flow convert to bubbly flow. Conversely, at high inlet oil velocities, stratified flow converts to dual continuous flow. Patterns and their evolution obtained with SPH are in accordance with the experimental patterns reported by [35] and with the flow pattern maps in the literature [36], [37]. This results prove the application of multiphase model in a popular engineering application getting good fitting between simulation and experiments.

# 4 Conclusions

In this work, a numerical multiphase model coupling a variable smoothing length and the shifting algorithm based on Smoothed Particle Hydrodynamics (SPH) has been developed and implemented using the DualSPHysics code. The multiphase model implemented in the DualSPHysics code improve the numerical results in the interface between two-fluids for multiphase SPH simulations.

For evaluating the multiphase implementation, several numerical validation test were conducted. The Poiseuille and Couette test cases, for two-fluids with different densities and viscosities were simulated and compared with the analytical solution. Results obtained for the multiphase model implemented in DualSPHysics provide a relative mean square error in the range $9.63 \times 10^{-7}$ to $3.16 \times 10^{-4}$. This result shows that our model that incorporate the shifting algorithm and the variable smoothing length formalism keeps good accuracy as compared with previous studies by [25], [26] and [27]. The accuracy of our model was evaluated through a convergence test where all validation cases were simulated for three different resolutions. As expected, accuracy clearly increases with resolution.

The root mean square error (RMSE) for the Poiseuille and Couette test cases with two densities are reported in Tables 2 and 4, respectively. The low RMSE error indicates good accuracy of the model for multiphase simulations that are usually affected by the presence of voids close to the areas where the interface is highly deformed. The coupling of the variable smoothing length formalism and the shifting algorithm prevents the creation of voids since it provides a better interface definition while keeping the continuity of the fluid.

The multiphase model coupling the shifting algorithm and the variable smoothing length formalism is able to better represent highly deformed interfaces and non-lineal effects in typical numerical examples of instabilities such as Rayleigh-Taylor and Kelvin-Helmholtz, as compared with other multiphase models reported by [30], [4] and [31].
The multiphase model is able to simulate properly the three forces that affect the dispersed phase in two-phase liquid-liquid flow; namely buoyancy, gravity and inertia force. The multiphase model coupling the variable smoothing length formalism and the shifting algorithm is able to generate numerical patterns of two phase flow for different superficial velocity ratios of fluids. Numerical results are comparable with real flows according to the flow pattern maps in the literature.

## Acknowledgment

The authors thank the financial support by the Mexican CONACyT, as well as ABACUS: Laboratory of Applied Mathematics and High-Performance Computing of the Mathematics Department of CINVESTAV-IPN. Our institution provided the facilities to accomplish this work.